\documentclass[pra,aps,superscriptaddress,twocolumn,notitlepage,showpacs]{revtex4-2}
\usepackage{latexsym}
\usepackage{amsmath}
\usepackage{amssymb}
\usepackage{graphicx}
\usepackage{caption}
\usepackage{subfigure}
\usepackage{float}
\usepackage{mathrsfs}
\usepackage{color}
\usepackage{txfonts}
\usepackage[justification=centering,
            format=plain]{caption}

\renewcommand{\raggedright}{\leftskip=0pt \rightskip=0pt plus 0cm}
\begin{document}

\title{Anti-interference Computational Ghost Imaging with Pink Noise Speckle Patterns}

\author{Xiaoyu Nie}
\affiliation{Department of Physics and Astronomy, Texas A\&M University, College Station, Texas, 77843, USA}
\affiliation{School of Physics, Xi'an Jiaotong University, Xi'an, Shaanxi 710049, China}

\author{Fan Yang}
\affiliation{Department of Physics and Astronomy, Texas A\&M University, College Station, Texas, 77843, USA}
\affiliation{School of Physics, Xi'an Jiaotong University, Xi'an, Shaanxi 710049, China}

\author{Xiangpei Liu}
\affiliation{Department of Physics and Astronomy, Texas A\&M University, College Station, Texas, 77843, USA}
\affiliation{Hefei National Laboratory for Physics Science at the the Microscale and Department of Modern Physics, University of Science and Technology of China, Hefei, Anhui 230026, China}

\author{Xingchen Zhao}
\affiliation{Department of Physics and Astronomy, Texas A\&M University, College Station, Texas, 77843, USA}

\author{Reed Nessler}
\affiliation{Department of Physics and Astronomy, Texas A\&M University, College Station, Texas, 77843, USA}

\author{Tao Peng}
\email{taopeng@tamu.edu}
\affiliation{Department of Physics and Astronomy, Texas A\&M University, College Station, Texas, 77843, USA}

\author{M. Suhail Zubairy}
\affiliation{Department of Physics and Astronomy, Texas A\&M University, College Station, Texas, 77843, USA}

\author{Marlan O. Scully}
\affiliation{Department of Physics and Astronomy, Texas A\&M University, College Station, Texas, 77843, USA}
\affiliation{Baylor Research and Innovation Collaborative, Baylor University, Waco, 76706, USA}
\affiliation{Princeton University, Princeton, New Jersey 08544, USA}

\date{\today}

\begin{abstract}
We propose a computational ghost imaging scheme using customized pink noise speckle pattern illumination. By modulating the spatial frequency amplitude of the speckles, we generate speckle patterns with a significant positive spatial correlation. We experimentally reconstruct images using our synthesized speckle patterns in the presence of a variety of noise sources and pattern distortion and shown it is robust to noise interference. The results are compared with the use of standard white noise speckle patterns. We show that our method gives good image qualities under different noise interference situations while the traditional way fails. The proposed scheme promises potential applications in underwater, dynamic, and moving target computational ghost imaging. 
\end{abstract}

\maketitle

\section{Introduction}
Ghost imaging (GI), which can be realized in both quantum and classical light~\cite{Pittman1995Optical,bennink2002two,valencia2005two, chen2009lensless}, is an alternative to the conventional image capture method using digital cameras. One major ameliorated system, computational ghost imaging (CGI), only employs one single-element detector to reconstruct images~\cite{shapiro2008computational}. The pre-determined speckle patterns for CGI are performed with spatial light modulators (SLM)~\cite{bromberg2009ghost}, digital micro-mirror devices (DMD)~\cite{radwell2014single}, LED arrays~\cite{xu20181000}, or optical phased arrays~\cite{li2018fast}. CGI also grants advantages in an expanding range of non-conventional applications such as wide spectrum imaging~\cite{edgar2015simultaneous,radwell2014single} and depth mapping~\cite{howland2013photon,sun2016single}. Moreover, CGI can be applied to images with spatially variant and re-configurable resolution~\cite{phillips2017adaptive,sun2018imaging,abetamann2013compressive}. 

By measuring the second-order correlation between the intensities of two light paths, thermal light GI can significantly eliminate interference from turbulence during the process of light propagation~\cite{meyers2011turbulence,shih2016turbulence}. Underwater CGI has also been demonstrated to attenuate the interference from the environment under certain conditions~\cite{le2017underwater}. To date, several studies such as differential detection~\cite{ferri2010differential,sun2013differential}, monitoring the noise~\cite{yang2018noise}, balanced detection~\cite{soldevila2016computational}, and micro-scanning techniques~\cite{zhao2017super,sun2016improving,sun2019improving} have been employed with CGI to decrease the influence of system noise further and enhance the signal-to-noise ratio (SNR). However, these methods are usually limited to a particular type of noise or require a large amount of extra work to eliminate the noise influence. On the other hand, orthogonal sampling strategies~\cite{luo2018orthonormalization,zhang2017hadamard}, compressive sensing ghost imaging~\cite{sun2017russian,katkovnik2012compressive,sheng2014correspondence}, and deep learning ghost imaging~\cite{lyu:2017aa} have been recently explored to obtain better image quality. These methods help shorten the signal acquisition time by reducing the total number of correlation measurements. However, these technologies rely upon pre-knowledge of the imaging system in advance. For example, compressive sensing ghost imaging needs prior understanding of the scene, such as sparsity constraints, to guide the image reconstruction. The deep learning ghost imaging method requires us to prepare thousands of training figures to develop convolutional neural networks. Improving the SNR of CGI without knowing image information, with general interference from background light noise, media scattering, pattern distortion,~\textit{etc.}, remains challenging. 

By modulating the phase at the Fourier plane, speckle patterns with desired probability density functions were achieved experimentally~\cite{bender2018customizing}. More recently, a sub-diffraction-limited resolution microscopy was demonstrated using such scheme~\cite{bender2021circumventing}. Different than modulating the phase front, we recently generated synthesize speckles via amplitude modulation of the spatial frequency of the input light, and achieve superresolving second-order correlation imaging with the obtained speckle illumination~\cite{li2021superresolving}. In this work, we adapt the pink noise concept to the spatial frequency domain of speckle patterns. Pink noise has been used to model electronic noise~\cite{dutta1981low} and the statistical structure of natural images~\cite{field1987relations}, and it is also one of the most common signals in biological systems~\cite{szendro2001pink}. We show a non-trivial positive correlation between a pixel and its neighbors in pink noise speckles. We then present a robust CGI scheme with the pink noise speckle patterns. The measurements are performed under several environmental interference situations. We also compare the results with the commonly used white noise speckle patterns. 

\section{Characteristics of pink noise speckles}
\begin{figure}[hbt!]
 \captionsetup{justification=raggedright,singlelinecheck=false}
	\includegraphics[width=\linewidth]{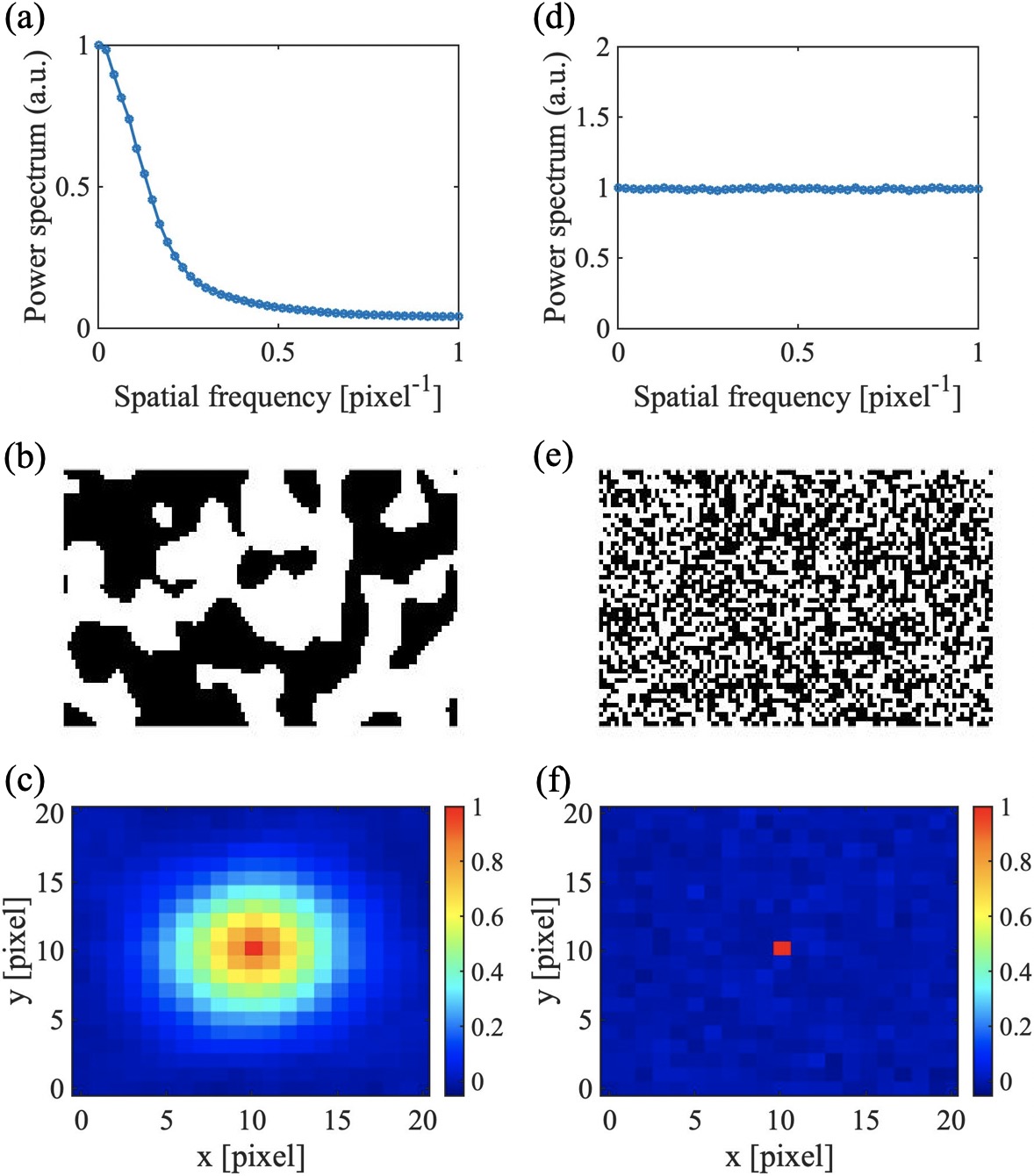}
	\caption{(a), (b) and (c): the 1D power spectrum, a typical speckle pattern, and correlation map of customized pink noise patterns; (d), (e) and (f): the 1D power spectrum, a typical speckle pattern, and correlation map of standard white noise patterns. Compared with white noise, there is a strong positive cross-correlation between a pink noise pixel and its neighbors.}
	\label{fig:red_white_comparison}
\end{figure}

Speckles for ghost imaging are generally produced by scattering laser light off a ground glass diffuser~\cite{martienssen1964coherence} or modulating the laser light using a spatial light modulator~\cite{bromberg2009ghost}. Here we introduce colored noise speckle patterns. The spectral power distribution of the speckles is $I(\omega)\simeq C_1 \delta (\omega) + C_2 \omega^n$ for spatial frequency $\omega$, where $C_1$ and $C_2$ are the coefficients of the \textit{DC} and colored noise spectrum components, respectively. For the standard white noise speckle patterns, $n=0$. For pink noise, we have $n=-1$, in which the spectrum decreases with spatial frequency. The spectrum distribution of pink noise and white noise used in the experiment are shown in Fig.~\ref{fig:red_white_comparison}(a) and (d), respectively. A random phase is assigned to each frequency component. The grayscale Gaussian pink and white noise patterns are then obtained via the inverse Fourier transform. Lastly we convert the patterns from grayscale to binary, which can be conveniently applied on the DMD later. The generated speckle patterns are shown in Fig.~\ref{fig:red_white_comparison}(b) and (e) for pink noise and white noise. The spectrum distributions maintain their desired distributions, \textit{i.e.}, pink noise and white noise spectra.   
Next, we examine the fluctuation correlation of the speckle patterns. To simplify the calculation without loss of generality, we consider here the one-dimensional case with positive frequencies.
The spatial intensity fluctuation correlation is defined as  
\begin{align}\label{fluctuation_correlation_defination}
\Gamma^{(2)} (\Delta x) &\equiv \langle I(x)I(x +\Delta x)\rangle-\langle I(x)\rangle \langle I(x +\Delta x)\rangle \cr
&=\mathcal{F}^{-1}\{|C_2 \omega^n|^2\}(\Delta x).
\end{align}

For white noise speckles ($C_2=C_{\mathrm{w}}$), there is no correlation between adjacent pixels
\begin{align}\label{white_fluctuation_correlation}
\Gamma^{(2)}_{\mathrm{w}} (\Delta x)=\mathcal{F}^{-1}\{|C_{\mathrm{w}}|^2\}(\Delta x) \propto \delta(\Delta x).
\end{align}
The pixelwise spatial correlation rapidly decays to zero, as shown in Fig.~\ref{fig:red_white_comparison}(f). 

For pink noise speckles ($C_2=C_{\mathrm{p}}$), we have the intensity fluctuation correlation as  
\begin{align}\label{pink_fluctuation_correlation}
\Gamma^{(2)}_{\mathrm{p}} (\Delta x)=\mathcal{F}^{-1}\{|C_{\mathrm{p}} \omega ^{-1}|^2\}(\Delta x).
\end{align}
If we examine the correlation with $\omega_1$ as the lowest frequency allowed which follows the pink poise spectral distribution~\cite{keshner19821}, and $\omega_2$ as the upper bound positive frequency used,  Eq.~(\ref{pink_fluctuation_correlation}) becomes 
\begin{align}\label{pink_fluctuation_correlation_2}
\Gamma^{(2)}_{\mathrm{p}} (\Delta x) &\propto \int^{\omega_2}_{\omega_1} |\omega^{-1}|^2 \cos (\omega \Delta x) d\omega \cr 
&=\frac{\cos(\omega_1 \Delta x)}{\omega_1}-\frac{\cos(\omega_2 \Delta x)}{\omega_2} \cr &+(\operatorname{Si} (\omega_1 \Delta x)- \operatorname{Si} (\omega_2 \Delta x) )\Delta x,
\end{align}
where $\operatorname{Si} (z) \equiv \int^{z}_0 \frac{\sin t}{t} d t$ is the sine integral. Such spatial frequency distribution leads to a remarkable positive cross-correlation between pixels adjacent to each other. This is in contrast to white noise patterns where there is no relation between different pixels, and the ensemble of fluctuation correlation is 0. To visualize this unique nature of pink noise, we randomly pick one pixel from pink and white noise patterns and calculate its fluctuation correlation with other pixels. We can see the striking difference in Fig.~\ref{fig:red_white_comparison}(c) for pink noise and Fig.~\ref{fig:red_white_comparison}(f) for white noise.

\begin{figure}[hbt!]
 \captionsetup{justification=raggedright,singlelinecheck=false}
	\includegraphics[width=0.95\linewidth]{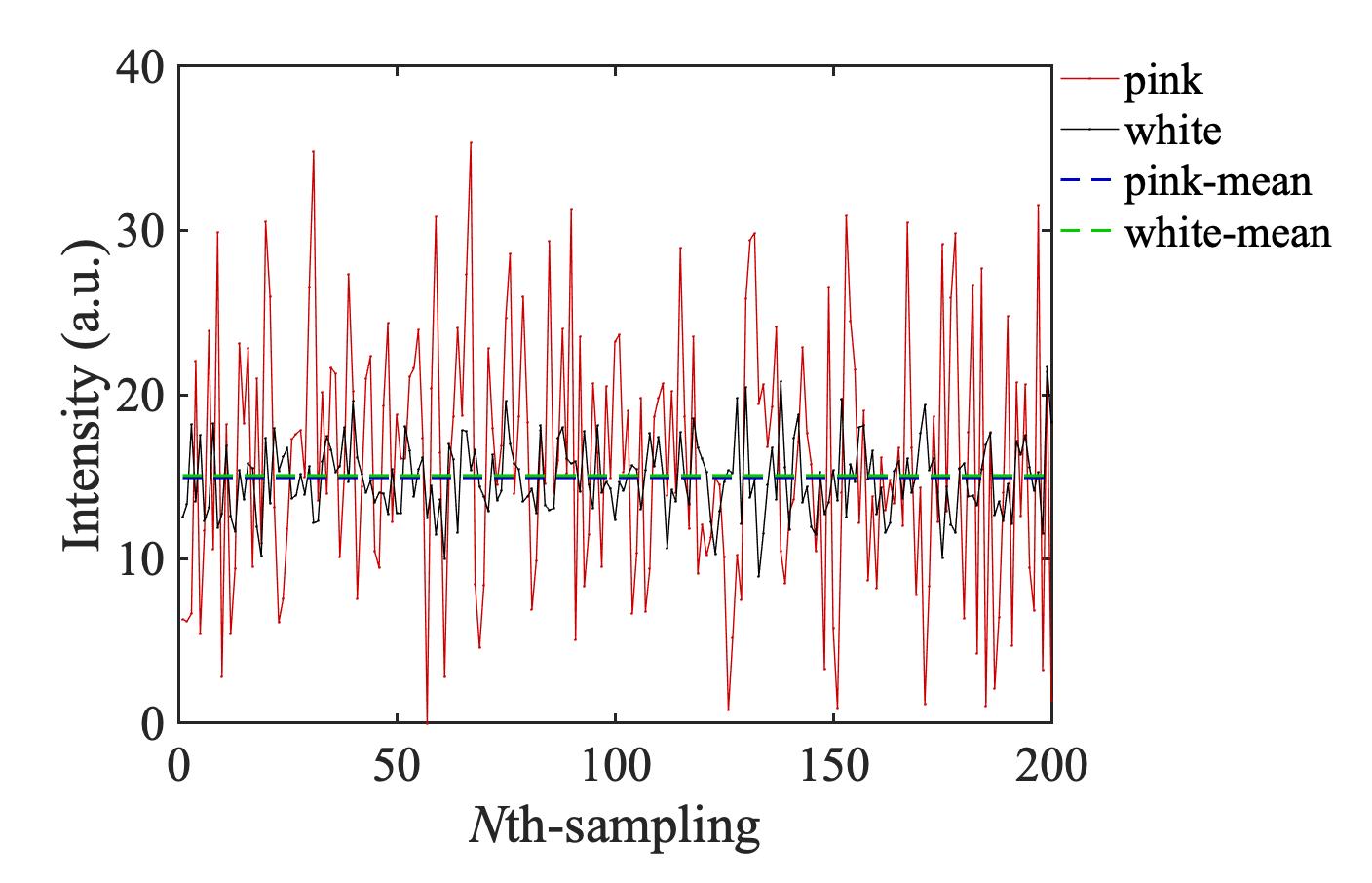}
\caption{Comparison of the intensity distribution for white and pink noise speckle patterns. A $10 \times 10$ pixel area was chosen and summed up to give the intensity distribution for a sequence of patterns. The pink noise speckle (in red) has a much larger fluctuation as compared to the white noise speckle (in black), while the average intensities (dashed lines) are the same.}
	\label{fig:intensity_distribution}
\end{figure}

In the imaging system, the second-order imaging is determined by the correlation function, for white noise: 
\begin{align}\label{white_noise_image}
\langle \Delta I_{\mathrm{b}}\Delta I_{\mathrm{w}}(x)\rangle \propto \langle\int d x_{\mathrm{o}} g^{(2)}_{\mathrm{w}}|T(x_{\mathrm{o}})|^2\rangle \approx |T(x_{\mathrm{o}})|^2,
\end{align}
where $I_{\mathrm{b}}$ is the bucket detector signal and $ I_{\mathrm{w}}(x)$ is the intensity of the white noise speckle at pixel $x$. $T(x_\mathrm{o})$ represents the object aperture function. 
The cross-correlation of the light on the image plane diffracted from different pixels has nearly zero contribution according to Eq.~(\ref{white_noise_image}).

Similarly, the second-order image measured with pink noise speckle pattern is given by
\begin{align}\label{pink_noise_image}
\langle \Delta I_{\mathrm{b}}\Delta I_{\mathrm{p}}(x)\rangle &\propto \langle\int d x_{\mathrm{o}} g^{(2)}_{\mathrm{p}}|T(x_{\mathrm{o}})|^2\rangle.
\end{align}
From Eq.~(\ref{pink_noise_image}), we notice that the situation is different due to the existence of cross-correlation between light from different speckles, as shown in Eq.~(\ref{pink_fluctuation_correlation_2}). Intuitively, we see that all the image pixels next to each other will contribute to cross-correlation with each pixel. This cross-correlation is in addition to the contribution of the auto-correlation from each pixel. The second-order signal strength is largely increased, and the noise is greatly suppressed due to the lack of correlation with other noises or the signals. 
To better view the advantage of using pink noise speckles vividly, we randomly pick an area of $10 \times 10$ pixels and sum them up as the bucket detector signal. We then plot the signal from a sequence of 200 patterns for both white and pink noise speckles, as shown in Fig.~\ref{fig:intensity_distribution}. We see that although the average intensities (in dashed lines) are almost the same, the fluctuations are significantly different. Given that the pink noise speckles have correlation with their neighborhoods, the much more significant fluctuations associated with the pink noise pattern suggest a much stronger fluctuation correlation between a single pixel (illuminating the object area) and the bucket signal in the CGI scheme. In the next section, we will experimentally show the advantage of pink noise speckle patterns in the CGI scheme, with the presents of a variety of strong interference.

\section{Experimental results}\label{sec:experiments}
\begin{figure}[hbt!]
 \captionsetup{justification=raggedright,singlelinecheck=false}
	\includegraphics[width=0.95\linewidth]{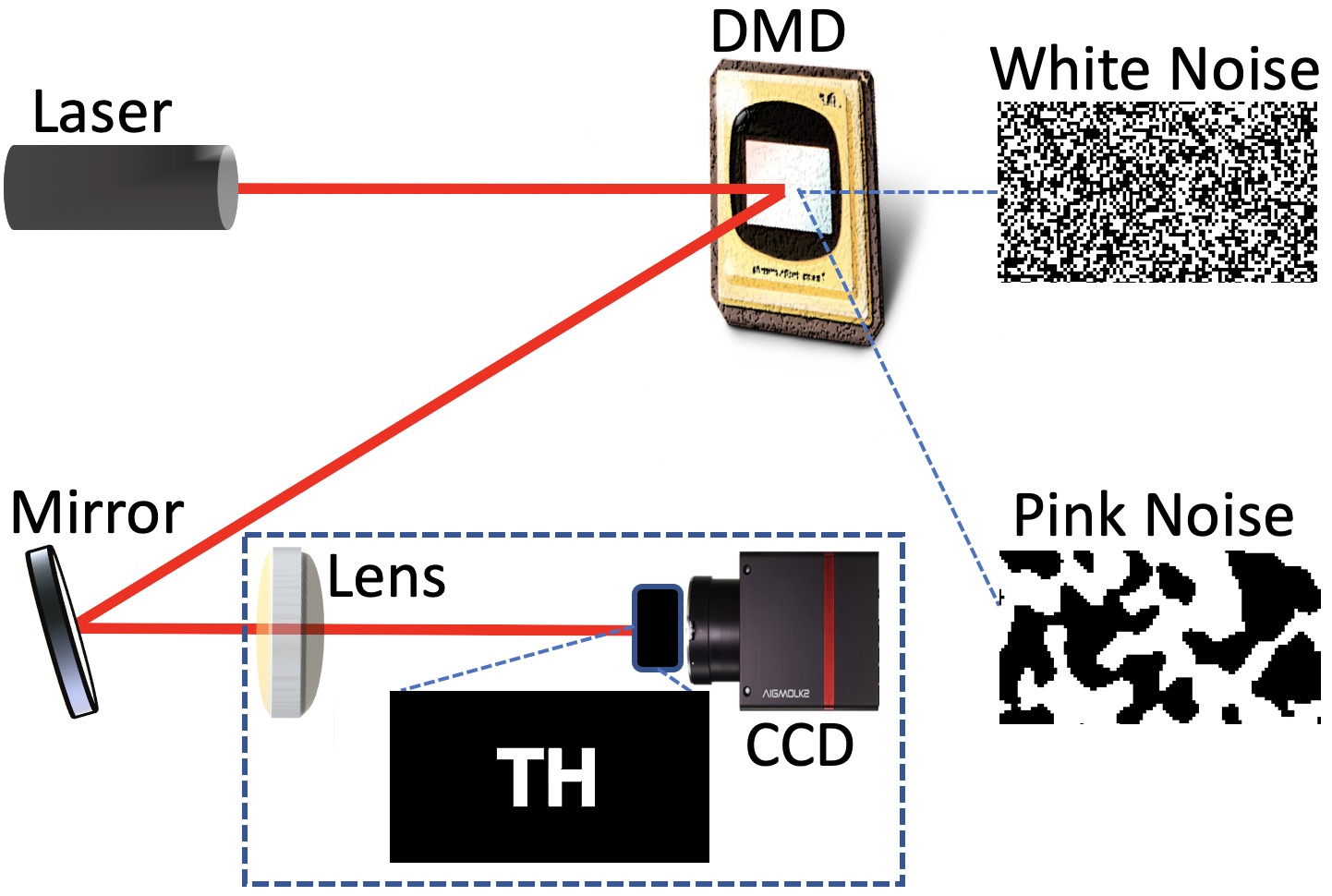}
	\caption{Schematic of the basic setup for a CGI experiment without interference. A CW laser is reflected by the DMD where the noise patterns are loaded. The reflected laser imprinted with noise patterns is imaged onto the object surface with the letters `TH'. A CCD put right after the object is used as a bucket detector in all the experiments in this work. The dashed frame is the part we modify by introducing a variety of noise sources.}
	\label{fig:setup}
\end{figure}

The experimental setup is shown in Fig.~\ref{fig:setup}. A CW laser is used to illuminate the DMD where the noise patterns are loaded. The pattern generated by the DMD is then imaged onto the object plane. A CCD right after the object is used as a bucket detector, \textit{i.e.}, only the total intensity on the CCD is used for the correlation measurement. In our experiment, the DMD contains tiny pixels (micro-mirrors), each measuring 16 $\mu$m $\times$ 16 $\mu$m. The noise pattern consists of 54 $\times$ 98 independent pixels (each pixel counts 4 $\times$ 4 DMD pixels). The object `TH' contains a total of about 600 independent pixels. In the following experiments, we introduce noise along the optical path between source and object, pattern distortion in addition to the optical path noise, noise on the detector, and pattern diffraction along the optical path. We perform CGI with both pink noise and white noise speckle patterns. To compare those two methods quantitatively, we introduce the SNR defined as
\begin{equation}
{\rm SNR} = \frac{\mu_{\mathrm{sig}}}{\sigma_{\mathrm{sig}}},
\end{equation}
where $\mu_{\mathrm{sig}}$ is the average signal value and $\sigma_{\mathrm{sig}}$ is the standard deviation of the signal. We show that for each case, the pink noise speckle pattern shows an anti-interference feature and gives a much higher SNR as compared to the white noise. 
\subsection{Noise between source and object}\label{sec:experiment1}

\begin{figure}[hbt!]
 \captionsetup{justification=raggedright,singlelinecheck=false}
	\includegraphics[width=0.9\linewidth]{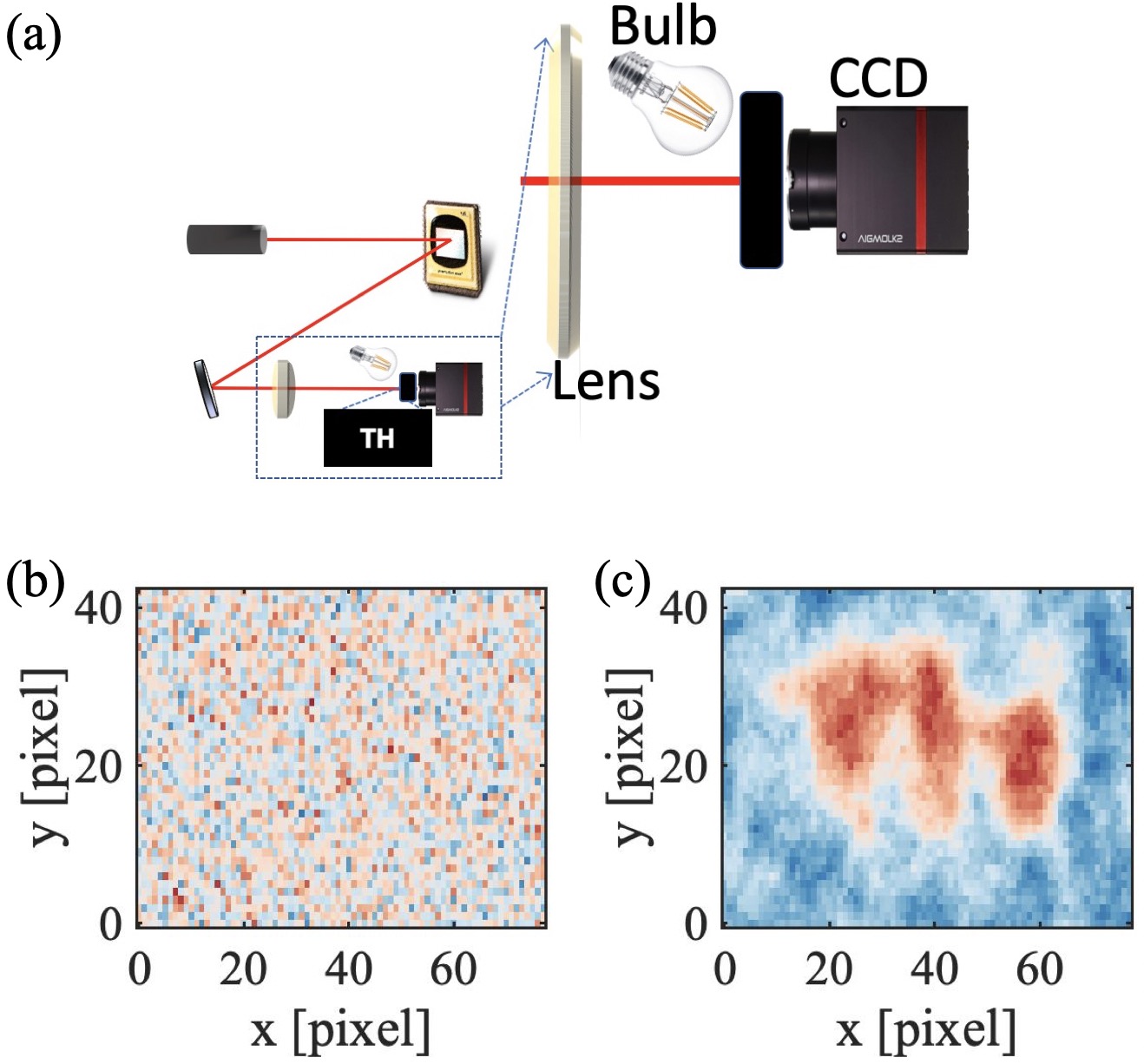}
	\caption{(a) Schematic of the setup with environmental noise introduced by a light bulb put in front of the object; (b) CGI with white noise speckle illumination using 800 patterns; (c) CGI with pink noise speckle illumination using 800 patterns. CGI with white noise is totally blurred due to the strong background noise, while the CGI with pink noise retrieves the image.}
	\label{fig:noise_before_object}
\end{figure}

The image quality of CGI depends largely on the SNR of the output intensity by the detector (the CCD in the present case). Therefore, a low noise level from both its own electronic noise and environmental noise is preferred. However, both noise sources exist in real applications. Here we use an incandescent lamp placed between DMD and the object to introduce a disturbance to the object's noise pattern illumination. The setup is shown in Fig.~\ref{fig:noise_before_object}(a). 800 white and pink noise speckle patterns are used in the measurements. The results are shown in Fig.~\ref{fig:noise_before_object}(b) and Fig.~\ref{fig:noise_before_object}(c), respectively. For the result of the pink noise CGI shown in Fig.~\ref{fig:noise_before_object}(c), the SNR is $\sim$ 4.40. The SNR of white noise CGI shown in Fig.~\ref{fig:noise_before_object}(b) is only $\sim$ 0.87. So when there is strong environmental noise along the optical path, in between the light source and the object, it is not easy to retrieve the image through a standard CGI scheme~\cite{le2017underwater}. It is nevertheless shown here that using pink noise speckles can suppress the influence of such disturbance to a great extent.

\subsection{Noise and diffuser between source and object}\label{sec:experiment2}
\begin{figure}[hbt!]
 \captionsetup{justification=raggedright,singlelinecheck=false}
	\includegraphics[width=0.9\linewidth]{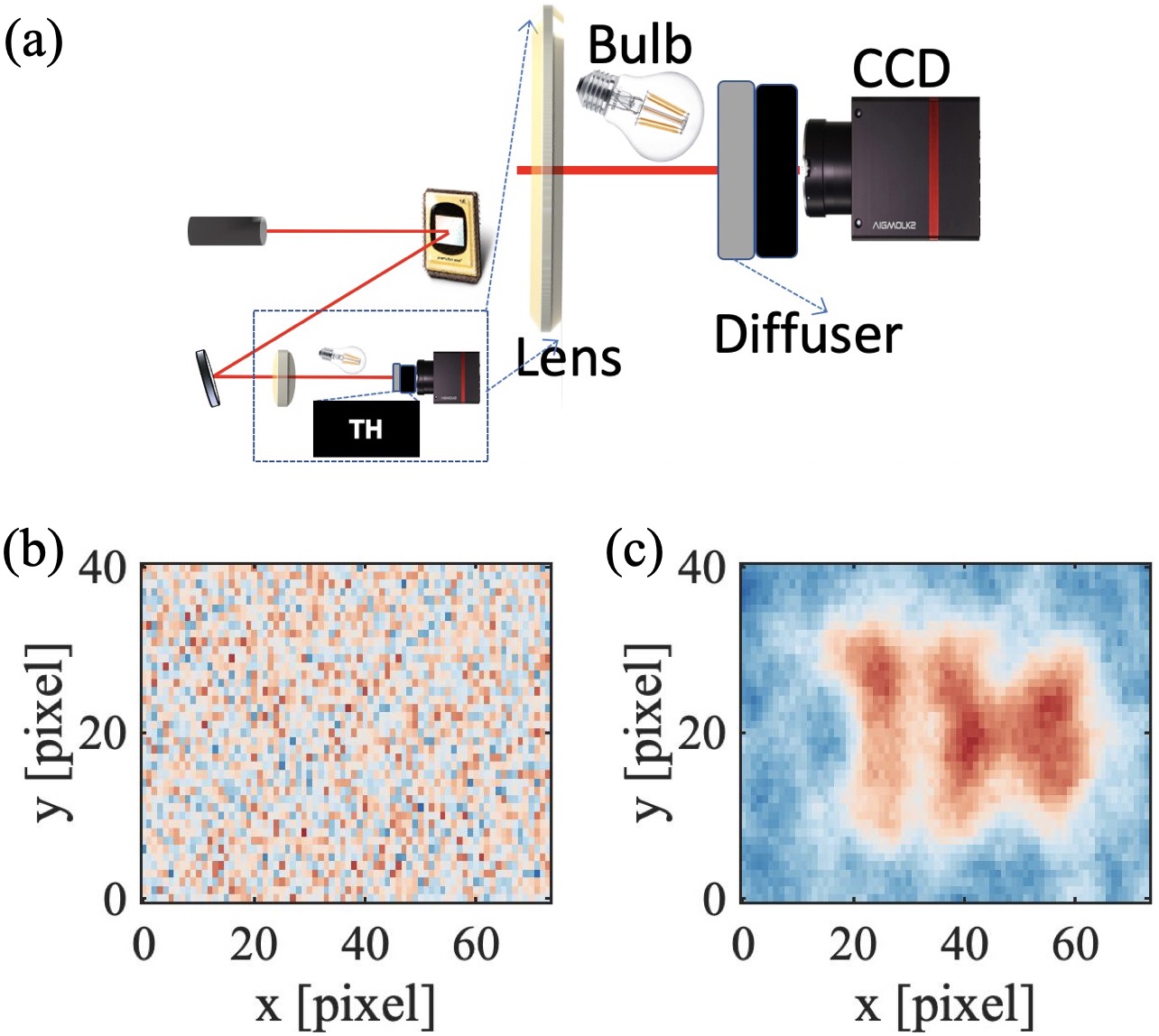}
	\caption{(a) Schematic of the setup. A ground glass diffuser (grit size 220) is put in front of the object to diffuse the speckle patterns, and a light bulb is also put in front of the object to introduce further environment noise; (b) CGI with white noise speckle illumination, a total of 800 patterns are used; (c) CGI with pink noise speckle illumination, a total of 800 patterns are used. CGI with white noise is totally blurred due to the strong background noise, while CGI with pink noise shows a retrieved image.}
	\label{fig:noise_plus_diffuser}
\end{figure}

In reality, the environment does not simply add noise along the optical path but also introduces turbulence and distortion. It has been shown that distortion along the optical path will greatly affect the image quality~\cite{le2017underwater}. In addition to the incandescent lamp, we add a ground glass diffuser between the lens and object to introduce diffraction and background noise simultaneously. This mimics the situation that the patterns are both smeared and buried in background noise. The schematic is shown in Fig.~\ref{fig:noise_plus_diffuser}(a). 
The CGI result of averaging 800 speckle patterns is shown in Fig.~\ref{fig:noise_plus_diffuser}(b) for white noise and Fig.~\ref{fig:noise_plus_diffuser}(c) for pink noise. The SNR of the white noise imaging is only 0.12, decreasing a lot compared to Sec.~\ref{sec:experiment1} (SNR $\sim$ 0.87). The SNR for pink noise is 4.13, comparable to the previous result (SNR $\sim$ 4.40). The introduction of the glass diffuser in the optical path decreases the image quality using white noise to a great extent, but it does not affect pink noise imaging much. This again is a demonstration of the robustness of pink noise CGI.

\subsection{Diffraction of speckle patterns}\label{sec:experiment4}
\begin{figure}[hbt!]
 \captionsetup{justification=raggedright,singlelinecheck=false}
	\includegraphics[width=0.9\linewidth]{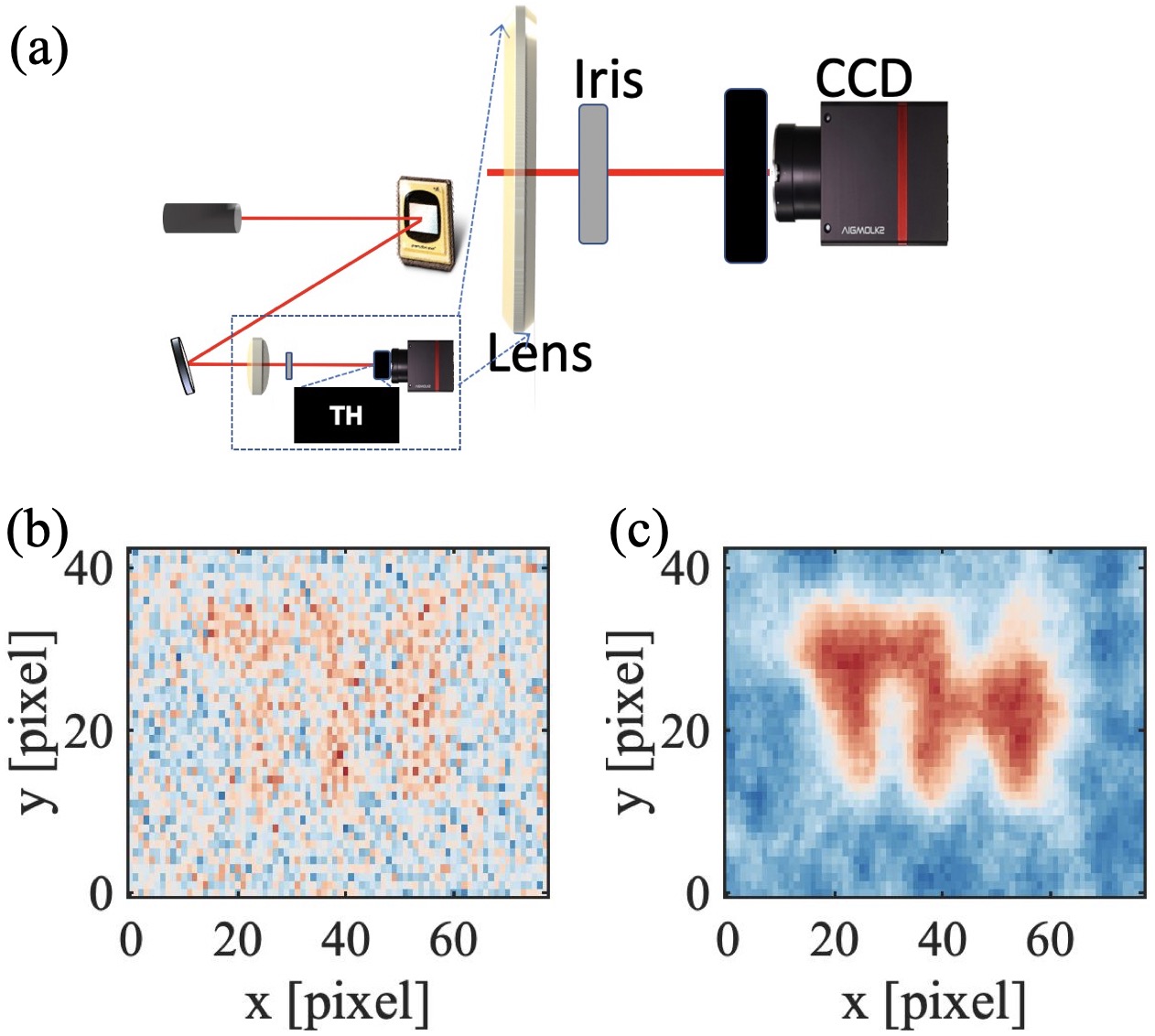}
	\caption{(a) Schematic of the setup with an iris inserted right after the lens. The diameter of the iris is 1000 $\mu$m. The speckle patterns are diffracted due to the iris; (b) CGI with the illumination of 800 white noise speckle patterns; (c) CGI with the illumination of 800 pink noise speckle patterns. CGI with white noise is totally blurred, while CGI with pink noise shows a clear retrieved image.}
	\label{fig:iris}
\end{figure}
In this part of our experiment, we put an iris right after the lens, which is used to image the speckle patterns on the object plane, as shown in Fig.~\ref{fig:iris}(a). In the iris's presence, the speckles can no longer maintain their spatial distribution as loaded on the DMD. Therefore the bucket detector recorded intensity is a mixture of desired speckles and unwanted speckles. The GI is expected to be destroyed since the one-to-one correspondence of the CGI is no longer valid. Indeed, the retrieved image from 800 white noise speckle patterns, as shown in Fig.~\ref{fig:iris}(b), is totally blurred. The SNR is only 0.88. On the other hand, pink noise CGI is still able to retrieve an image of the object with 800 patterns, as shown in Fig.~\ref{fig:iris}(c), and the SNR is 3.75. 

\subsection{Noise between object and detector}\label{sec:experiment3}

\begin{figure}[hbt!]
 \captionsetup{justification=raggedright,singlelinecheck=false}
	\includegraphics[width=0.9\linewidth]{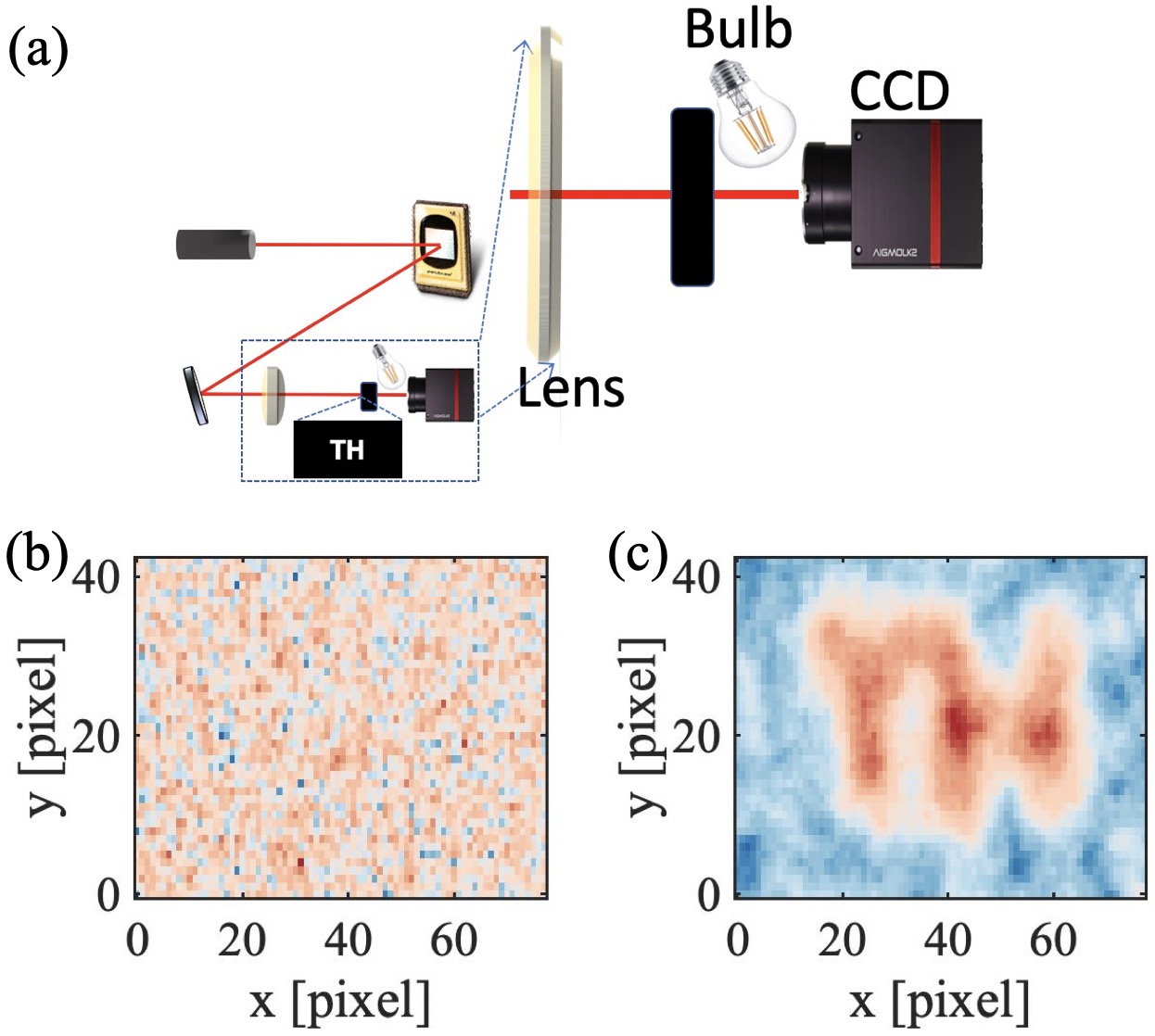}
	\caption{(a) Schematic of the setup with environmental noise introduced by a light bulb put in front of the detector; (b) CGI with white noise speckle illumination, 10000 patterns used for the ensemble average; (c) CGI with pink noise speckle illumination, 10000 patterns used for the ensemble average. CGI with white noise is totally blurred due to the strong background noise, while CGI with pink noise shows a clear retrieved image.}
	\label{fig:noise_after_object}
\end{figure}

To further demonstrate the strength of pink noise CGI, we enhance the interference level by placing an incandescent lamp that produces strong light noise between the object and the CCD. Under certain circumstances, such as in biomedical applications, the signal is weak due to significant attenuation and diffusion along the optical path. In such cases, the signal at the detector could be below the detector noise level. Here we show that even in those extreme situations, our scheme can still retrieve the object image. As shown in Fig.~\ref{fig:noise_after_object}(a), the lamp introduces noise distributed uniformly on the CCD plane to mimic strong noise from the bucket detector. To be more specific, the noise intensity of each pixel is around 130 to 140 units, whereas the transmitted signal is only around 2 to 3 units. The experimental results are presented in Fig.~\ref{fig:noise_after_object}(b) for white noise illumination and Fig.~\ref{fig:noise_after_object}(c) for pink noise illumination. Here due to the extreme background noise at the detector, 10000 patterns are used to retrieve the image in both cases. As expected, the white noise speckle patterns fail to reconstruct the image at all, and the SNR is only 0.19. However, the pink noise in CGI can retrieve the image of the object, with an SNR of $\sim$ 3.97. 


\section{Summary}
We have developed a novel method to create the pink noise speckle pattern and applied it to the CGI system. The modulation on the spectrum domain enables us to create speckle patterns that have strong positive fluctuation correlation between pixels. This feature makes it robust to noisy environment in the CGI system. The anti-interference feature of the pink noise CGI is experimentally demonstrated. We examined and compared the SNR of images retrieved by pink noise speckle patterns and standard white noise patterns. Four types of noisy environments are introduced to mimic the random noise and pattern distortion along the optical path and the shallow signal to noise ratio at the bucket detector. We have shown that the resulting SNR of pink noise CGI is always much better than that of white noise CGI in the presence of different noise sources and interference from obstacles. This work is of great significance for the practical application of CGI due to its robustness.

\section*{Acknowledgement} 
X. N. and F. Y. thank A. Svidzinsky for his kind help during their visit to IQSE, Texas A\&M University. We gratefully acknowledge the support of Air Force Office of Scientific Research (Award No. FA9550-20-1-0366 DEF), Office of Naval Research (Award No. N00014-20-1-2184), Robert A. Welch Foundation (Grant No. A-1261), National Science Foundation (Grant No. PHY-2013771), and Qatar National Research Fund (project NPRP 13S-0205-200258).

X. N. and F. Y. contributed equally to this work.

\end{document}